# Dynamical Scattering in Coherent Hard X-Ray Nanobeam Bragg Diffraction

A. Pateras,[1] J. Park,[1] Y. Ahn,[1] J. A. Tilka,[1] M. V. Holt,[2] H. Kim,[3] L. J. Mawst,[3] and P. G. Evans[1]

[1] *Department of Materials Science & Engineering, University of Wisconsin-Madison, Madison, Wisconsin 53706 USA*

[2] *Center for Nanoscale Materials, Argonne National Laboratory, Argonne, IL 60439, USA*

[3] *Department of Electrical and Computer Engineering, University of Wisconsin-Madison, Madison, WI 53706, USA*

Unique intensity features arising from dynamical diffraction arise in coherent x-ray nanobeam diffraction patterns of crystals having thicknesses larger than the x-ray extinction depth or exhibiting combinations of nanoscale and mesoscale features. We demonstrate that dynamical scattering effects can be accurately predicted using an optical model combined with the Darwin theory of dynamical x-ray diffraction. The model includes the highly divergent coherent x-ray nanobeams produced by Fresnel zone plate focusing optics and accounts for primary extinction, multiple scattering, and absorption. The simulation accurately reproduces the dynamical scattering features of experimental diffraction patterns acquired from a GaAs/AlGaAs epitaxial heterostructure on a GaAs (001) substrate.

Synchrotron radiation light sources and x-ray free electron lasers produce bright coherent x-ray nanobeams that enable the implementation of coherent x-ray diffraction imaging (CXDI) techniques allowing the three-dimensional (3D) visualization of strain and other crystallographic features [1-5]. In CXDI and Bragg ptychography [6,7], the formation of real-space images that contain the lattice displacement relies on the use of the forward and inverse Fourier transform in iterative algorithms under the kinematical approximation [8]. The kinematical approximation is valid for small crystals, in which primary extinction, multiple scattering, and photoelectric absorption can be neglected [9,10]. In larger crystalline systems such as multilayer oxide heterostructures, far-from-surface optically active defects in quantum materials, strain-engineered semiconductors, and epitaxial bandgap-engineered quantum wells and quantum dot structures, however, the kinematical theory does not allow neither the visualization of strain through phase retrieval nor the quantitative prediction of x-ray diffraction patterns. Removing this limitation is a key step to quantitatively image strain through phase retrieval techniques. We report the observation of dynamical effects in nanobeam diffraction and the development of a simulation method that employs the dynamical theory of x-ray diffraction combined with a wave-optics model. The consideration of dynamical diffraction effects allows nanobeam diffraction to be extended into a new regime, including systems consisting of crystals or sublayers of crystals that have sizes exceeding the x-ray extinction depth (i.e. at the micron scale for hard x-rays) or when nanoscale features are formed on bulk single-crystal surfaces. A comparison with coherent x-ray nanobeam diffraction experiments illustrates that dynamical effects in the nanobeam diffraction patterns of GaAs heterostructures can be accurately described and reproduced by this optical simulation approach. The results expand the applicability of nanobeam diffraction methods to complex layered crystals and point to further directions in accounting for dynamical effects in coherent diffraction simulations in x-ray coherent diffraction imaging and ptychography.

A straightforward comparison of dynamical and kinematical diffraction demonstrates the importance of dynamical effects in x-ray nanobeam and coherent diffraction studies of nanoscale crystals. Recently, CXDI studies of nanoparticles with sizes approaching or greater than the x-ray extinction depth show how dynamical diffraction effects corrupt the reconstructed images from phase retrieval algorithms [11]. The conditions under which dynamical diffraction artifacts can be neglected in the case of finite size nanoparticles and the impact of using datasets generated from different theoretical models on the final reconstructed images have been quantified [12]. Fig. 1 shows two examples that illustrate the bounds of the applicability of coherent diffraction simulations presently underpinned by the kinematical approximation. Fig. 1 compares dynamical and kinematical descriptions of the diffraction of a $\pi$-polarized x-ray plane wave with a photon energy of 10.4 keV for the 004 x-ray reflection of a finite-thickness GaAs crystal. The extinction depth under these conditions is 0.69 μm, corresponding to $1.2 \times 10^3$ GaAs unit cells [9]. Fig. 1(a) shows the full-width at half-maximum (FWHM) angular widths predicted using the kinematical approximation lattice sum and the Darwin dynamical theory [13,14]. The results of both methods are equivalent in the small-thickness regime. For crystals with thicknesses greater than the x-ray extinction depth, however, the angular width predicted by the kinematical theory differs significantly from the more accurate dynamical theory. The dynamical FWHM saturates at a value a factor of 1.14 larger than the Darwin width for thick crystals, resulting from the different definitions of the Darwin and FWHM widths. The kinematical theory, in comparison, predicts a decrease in the angular width without a lower bound. Fig. 1(b) shows the normalized kinematical and dynamical theory peak intensity reflectivity. The kinematical theory predicts a non-physical divergence of the diffracted intensity from thick crystals, exceeding unity, and not matching the saturation predicted by the dynamical theory [10,15]. These straightforward effects illustrate the need to consider dynamical diffraction in systems with crystal sizes greater than approximately 500 nm.

Figure 2 illustrates the key differences between diffracted patterns predicted using the kinematical and dynamical theories. Figures 2(a) and (b) show the predicted x-ray intensity of the 004 Bragg reflection using the same structural parameters for the quantum well heterostructure, employing the kinematical and dynamical approaches with a plane-wave incident beam. The kinematical theory predicts a non-physically high intensity for the substrate peak, does not account for the shift in reflections due to refraction, and does not correctly include the interference of the substrate reflection with diffraction from the thin film. In addition, the high-frequency oscillations originating from the substrate thickness are never observed in reality due to the high absorption away from the region of total reflectivity. The dynamical theory prediction shown in Figure 2(b) correctly accounts for all of these effects and accurately reflects what is seen in experiments. An important feature of the dynamical prediction is the angular range of near-unity reflectivity as experimentally observed, known as the Darwin width. The dynamical theory is particularly important in predicting the diffraction patterns of samples with thick crystalline layers, including single-crystal substrates, or in cases where the diffracted amplitude from thin layers falls in the same angular range as the diffraction from thick layers. In the sample considered here, the only dynamically diffracting layer is the GaAs substrate. In other systems, however, thick crystalline epitaxial layers such as semiconductor superlattices can require the use of dynamical techniques in order to simulate the diffracted intensity accurately.

The problem considered in this manuscript is how to extend the previous description of nanobeam diffraction to include important dynamical diffraction effects. The nanobeam diffraction simulation presented here employs the Darwin theory of dynamical diffraction, accounting for primary extinction, multiple scattering, absorption, and refraction. Other dynamical effects, such as many-beam diffraction, lateral transport of x-ray intensity parallel to the surface, and lateral inhomogeneity within the illuminated volume are not described in the Darwin approach, and thus, are not part of the present work [10,15,16]. A key validation of this modeling approach is provided by comparing it in detail with a nanobeam diffraction experiment. Diffraction patterns acquired from a GaAs/AlGaAs heterostructure exhibit narrow and intense x-ray reflections and scattered x-ray intensity at angles outside the nominal divergence angle of the focused radiation.

The GaAs/AlGaAs heterostructure was epitaxially grown on a 450 µm-thick, (001)-oriented GaAs substrate by metalorganic vapor deposition. As illustrated in Fig. 3(a), the heterostructure consists of a 14.5 nm-thick GaAs layer between two 115 nm-thick $Al_{0.24}Ga_{0.76}As$ layers. These layers have lattice parameters, $a_{GaAs}$=5.65325 Å and $a_{AlGaAs}$=5.65573 Å, giving an epitaxial lattice mismatch $\varepsilon=a_{AlGaAs}/a_{GaAs}-1$ of $4.4\times10^{-4}$ [17]. Precise thicknesses in the nanoscale region probed in the x-ray nanobeam experiments were obtained by comparing the experimental data with the simulation result, as described below.

Nanobeam dynamical diffraction experiments were conducted at the Hard X-ray Nanoprobe at the Advanced Photon Source, using the x-ray focusing optics and scattering geometry shown in Fig. 3(b). An x-ray beam with photon energy of 10.4 keV was prepared using a two-bounce Si (111) monochromator and was focused to a spot with 30 nm nominal FWHM with a Fresnel zone plate (FZP). The FZP had a diameter of 150 µm and outermost zone width of 20 nm, leading to a focal distance of 26.8 mm and an angular divergence $\delta$=0.34°. The unfocused order of the beam was blocked by a 60 µm-diameter center stop (CS). Radiation focused to higher orders was eliminated by a 30 µm-diameter order-sorting aperture (OSA). The incident x-ray beam is linearly polarized with the vector of the electric field in the horizontal plane of the storage ring. The sample is set in a horizontal scattering geometry, making the x-ray electric field nearly parallel to the scattering plane. The incident beam is thus considered to be π-polarized in the simulation. The effective angle of incidence θ is defined as the angle between the surface of the sample and the center of the focused x-ray beam, as depicted in Fig. 3(b).

The diffracted beam intensity was recorded using a pixel-array detector (Pixirad-1, PIXIRAD Imaging Counters s.r.l.) located 0.9 m from the sample. The two-dimensional diffraction pattern is spanned by angles 2θ and χ, defined as the angle lying on the horizontal plane, and the direction normal to the beam footprint direction, respectively. The detector pixels are arranged hexagonally with pixel centers spaced by 52 μm and 60 μm along the directions spanned by the angles 2θ and χ, respectively [18]. The analysis below treats the grid of pixels as rectangular and neglects the offset in 2θ between centers of rows displaced by a single pixel along χ. The acquisition times for each diffraction pattern ranged from 0.2 to 0.5 sec, depending on the selection of the filters that were used to attenuate the incident beam to avoid saturating the detector with the bright GaAs 004 substrate reflection.

A nanobeam diffraction pattern obtained at the nominal Bragg angle for the GaAs 004 reflection is shown in Fig. 4(a). The angular divergence introduced by the FZP results in a wide angular distribution of intensity in the diffraction pattern. The bright sharp vertical line of x-ray intensity in the center of the diffraction pattern arises from diffraction from the 450 μm-thick GaAs substrate and cannot be reproduced by the kinematical theory. This sharp intensity feature is a key illustration of the importance of incorporating dynamical diffraction effects into coherent x-ray nanobeam diffraction models. Other features and dynamical effects are discussed in detail below.

The nanobeam dynamical diffraction simulation consists of propagating the x-ray wavefield through focusing optics, to the sample, and a combined description of the diffraction from the sample. The x-ray wavefield at the focus was calculated using an optical simulation method in which the x-ray wavefield incident on the FZP is assumed to be a monochromatic plane wave [19]. The FZP imprints a phase on the incident wavefront and the CS and OSA are implemented in this approach as objects with complex dielectric constants, largely serving to attenuate unfocused radiation and radiation focused to higher orders of foci [19,20]. CXDI and ptychography techniques cannot yet be used to retrieve the focal spot intensity and phase in closely lattice-matched epitaxial heterostructures or materials systems incorporating large crystals and we have thus used a more idealized optical description of the focused beam.

The calculation of the dynamical reflectivity of the heterostructure includes multiple layers with different compositions and lattice parameters [14]. The method reported in Ref. [14] considers the case of a perfectly collimated plane wave, incident at a single angle of incidence $\theta_i$. Crystallographic unit cells are numbered from the bottom layer at the interface with the substrate ($k = -1$), to the surface ($k = -N$). The reflectivity at the unit cell with index $k$ is a function of the scattering vector $Q_z = \frac{4\pi}{\lambda} \sin \theta_i$:

$$r_k(Q_z) = -ig + \frac{(1-ig_0)^2}{ig + e^{-2i\phi} r_{k+1}^{-1}(Q_z)} \qquad (1).$$

Here $g$ is the amplitude reflectivity of a single unit cell, $g_0$ is the amplitude of the forward-scattered wavefield from a single unit cell, $\phi = \frac{2\pi}{\lambda} a \sin \theta$ is the phase shift of the forward-scattered wavefield resulting from propagation through a single-unit-cell thickness, $\lambda$ is the x-ray wavelength, and $a$ is the lattice parameter. The dependence of the structure factor on the composition of each layer is included in $g$ and $g_0$ [9,14], which also depend on the layer index $k$. The values of $g$ and $g_0$ are given by [21]:

$$g = \frac{\lambda r_e}{\sin\theta_i} MF(\theta_i) \cos 2\theta_i \qquad (2a)$$

and
$$g_0 = \frac{\lambda r_e}{\sin\theta_i} MF(\theta_i=0) \qquad (2b).$$

Here, $r_e$ is the classical electron radius, $M$ is the number of unit cells per unit area, and $F(\theta)$ is the unit-cell structure factor, which depends on the composition of the unit cell for which it is computed. Equations (2a) and (2b) include the polarization factor $\cos 2\theta_i$ to account for the π-polarized incident beam in our experiments. The case of σ-polarization can be considered by replacing $\cos 2\theta_i$ in Eq. (2a) with unity, giving $g = \frac{\lambda r_e}{\sin\theta_i} MF(\theta_i)$. The factors $g$ and $g_0$ are complex, thus, allowing absorption effects to be included. Note that the dynamical reflectivity is a unitless complex-valued quantity with a magnitude equal to, or less than one.

The dynamical reflectivity from the thin-film heterostructure is calculated recursively, using the known reflectivity of the substrate to initialize the calculation in Eq. (1). The substrate reflectivity $r_0$ is given by the Darwin-Prins reflectivity [10]. Alternatively, initializing the calculation with $r_0=0$, would correspond to the case in which there is no contribution from the substrate. The simulation method can further be extended to include strained epitaxial layers or the effects of strain gradients by varying the lattice parameters used in the simulation.

The dynamical reflectivity was evaluated at a series of values of $Q_z$, spanning the range of incident wavevectors in the x-ray wavefield at the focal spot [13]. In discretized form, the focused x-ray wavefield is a complex-valued two-dimensional matrix with values assigned at every real-space pixel of the focal plane. The reciprocal space range of wavevectors composing the incident wavefield was determined by computing the two-dimensional discrete Fourier transform of the focused-beam wavefield. The diffracted intensity at the detector plane was obtained by calculating the absolute square of the element-wise product of the focused beam propagated to the far-field with the dynamical reflectivity at the corresponding wavevector. The diffracted intensity is expressed in the coordinate frame of the detector using the angular transformation given in Ref. [13].

A comparison of experimental and simulated diffraction patterns for the GaAs/AlGaAs heterostructure is shown in Figs. 4(a) and 4(b). The diffraction patterns are shown for an effective incidence angle $\theta$ that satisfies the GaAs (004) Bragg condition. The common key features observed in both experimental and simulated diffraction patterns are (i) a bright, sharp feature arising from the GaAs (004) reflection in the horizontal center of the diffraction pattern and (ii) broader features of lower intensity originating from the GaAs quantum well layer and the two AlGaAs layers. Due to the extremely small lattice mismatch, the diffracted signal from the three thin layers of GaAs and AlGaAs overlaps with the signal from the substrate. The intensity distribution from the GaAs/AlGaAs layers is centered at a value of $2\theta$ 0.01° less than the substrate reflection, as predicted by the composition of the AlGaAs. The less intense vertical lines of intensity in Fig. 4(a) correspond to thickness fringes from the three thin layers. The maximum intensity of these fringes is three to four orders of magnitude lower than the peak intensity of the substrate reflection and is lower at angles far from the substrate reflection. The fringes with highest contrast have angular spacing $\delta_{2\theta} = 0.033°$, corresponding to a thickness $t$=114 nm, obtained using $t=\lambda/(2\ cos\theta\ \delta_{2\theta})$ [22], in agreement with the value of 115 nm employed in the simulation. Each of these fringes is also split due to interference between the two AlGaAs layers. The signal of the quantum well layer cannot be observed as a readily separate intensity maximum in the diffraction pattern because of the small lattice-mismatch.

Fig. 4(c) shows a comparison of experimental and simulated diffraction patterns integrated along the angle $\chi$. In order to account for a small experimental uncertainty in the absolute value of 2θ, integrated experimental and simulated intensities are plotted as a function of the angular difference $2\theta-2\theta_{center}$, where $2\theta_{center}$ is the center of the predicted or measured distribution of diffracted intensity. The values of the layer thicknesses of the heterostructure were determined by examining the angular positions of the thickness fringes originating from the GaAs and AlGaAs layers. The optimum agreement between simulation and experiment was obtained with layer thicknesses of 115 nm AlGaAs/14.5 nm GaAs/115 nm AlGaAs. A kinematical scattering model of the nanobeam diffraction would provide a poor match for the experimental result in Fig. 4(a) because the strong peak and tails of the scattering from the GaAs substrate would be incorrectly described in the kinematical approximation.

The intensity distributions in simulated and experimental diffraction patterns are compared for a wide range of x-ray incident angles θ in Fig. 5. Diffraction patterns were acquired by scanning θ from 24.75° to 25.15°, while moving the detector with steps of twice the size of the sample rotation. For a collimated incident beam, the scan would correspond to a conventional θ/2θ scan in which the truncation rod of intensity from the substrate is tracked by a point detector.

Several features of Fig. 5 arise from dynamical diffraction effects that cannot be reproduced by the kinematical theory. The bright substrate peak and the positions and profile of the thickness fringes due to the interference of the substrate reflection with diffraction from the thin film cannot be quantitatively reproduced by the kinematical theory. The simulated and experimental diffraction patterns acquired at an angle of 24.75°, below the nominal Bragg angle, exhibit a bright sharp line of intensity which appears outside the angular range corresponding to the divergence of the focused x-ray beam. The sharp line observed on the high-2θ edge of the diffraction pattern arises from the very strong reflection of the angular intensity tails of the focused x-ray nanobeam by the substrate. As the angle of incidence increases, the line of intensity arising from the substrate enters the geometric cone of the focused radiation and produces a strong line of intensity. A comparison of the images in Figs. 5(a)-(h) shows that the simulation reproduces the features on the experimental diffraction patterns, including the angular divergence of the focused beam, the shadow of the CS, and the uniform distribution of illumination within the focused beam.

Several differences between simulation and experiment in Fig. 5 arise from the difference between the idealized description of the x-ray source and focusing optics in the modeling and the more complex situation in the experiment. Concentric intensity appearing in the experimental diffraction pattern are due to artifacts in the FZP fabrication, which have been previously visualized using x-ray ptychography and electron microscopy [23,24]. In addition, the zone-doubling technique used to fabricate the hard x-ray optic capable of focusing hard x-ray beams to an intensity FWHM of tens of nm consists of lithographically overlaying complex engineered structures with smallest feature size of approximately 50 nm, sometimes resulting in the observation of wavefront aberrations that would not be present in an idealized optic with 20 nm outermost zones [24]. In our experimental case this resulted in a small halo visible around the principal cone of the optic which is not comparable to the main intensity pattern except on the sensitive logarithmic scale employed for imaging and does not impact our results. Finally, the finite energy bandwidth of the Si (111) monochromator causes the experimentally measured width of the substrate reflection to exceed the predicted dynamical diffraction Darwin width. The angular Darwin width of the GaAs 004 reflection is given by $2g/m\pi \tan\theta = 1.6$ mdeg where $m=4$ and $g$ is given by Eq. (2a) [9]. The energy bandwidth of the photon energy $E$ of the incident x-ray beam focused x-ray beam is $\Delta E/E = 1.3 \times 10^{-4}$, which gives rise to an angular difference of 3.5 mdeg between diffracted beams excited by different photon energies. The experimentally observed angular width is approximately two pixels, corresponding to 6 mdeg, consistent with the expected broadening due to the energy bandwidth of the monochromator.

By including dynamical diffraction effects in x-ray nanobeam diffraction it is possible to predict and numerically reproduce the diffraction patterns from arbitrary heterostructures consisting of thin layers epitaxially grown on thick single-crystal substrates. Beyond the GaAs/AlGaAs heterostructure described here, dynamical scattering simulation methods have the potential to extend the capabilities of CXDI methods. Dynamical effects will be particularly important in CXDI studies of crystals with thicknesses larger than the x-ray extinction depth and in systems combining large crystals with nanoscale features. In such cases, artifacts appear in the retrieved amplitude and phase of CXDI reconstructions due to primary extinction, refraction, and absorption and the artifacts can be treated by applying corrections to the retrieved image based on the Takagi-Taupin equations [11,12]. It has not yet been possible, however, to incorporate dynamical diffraction effects directly in a phase-retrieval algorithm to account for primary extinction, multiple scattering, refraction, and absorption. The approach described here represents a step towards the solution of the phase problem if a dynamical simulation can be introduced within an optimization algorithm for phase-retrieval. Integrating dynamical diffraction with nanobeam diffraction, CXDI, and Bragg ptychography has the potential to allow the visualization of strain and defects relevant to key scientific problems in electronics, magnetic materials, and materials for energy applications.

A.P., J.P., Y.A., and P.G.E. were supported by the U.S. DOE, Basic Energy Sciences, Materials Sciences and Engineering, under Contract No. DE-FG02-04ER46147 for the x-ray scattering studies and analysis. J.A.T. acknowledges support from the National Science Foundation Graduate Research Fellowship Program under Grant No. DGE-1256259. Use of the Center for Nanoscale Materials and the Advanced Photon Source, both Office of Science user facilities, was supported by the U.S. Department of Energy, Office of Science, Office of Basic Energy Sciences, under Contract No. DE-AC02-06CH11357. Laboratory characterization at the University of Wisconsin-Madison used instrumentation supported by the National Science Foundation through the UW-Madison Materials Research Science and Engineering Center (DMR-1121288 and DMR-1720415).

[1]     I. K. Robinson, I. A. Vartanyants, G. J. Williams, M. A. Pfeifer, and J. A. Pitney, Physical Review Letters **87**, 195505 (2001).

[2]     P. Thibault, M. Dierolf, A. Menzel, O. Bunk, C. David, and F. Pfeiffer, Science **321**, 379 (2008).

[3]     S. Brauer, G. B. Stephenson, M. Sutton, R. Bruning, E. Dufresne, S. G. J. Mochrie, G. Grubel, J. Alsnielsen, and D. L. Abernathy, Physical Review Letters **74**, 2010 (1995).

[4]     W. Yun, B. Lai, Z. Cai, J. Maser, D. Legnini, E. Gluskin, Z. Chen, A. A. Krasnoperova, Y. Vladimirsky, F. Cerrina, E. Di Fabrizio, and M. Gentili, Rev. Sci. Instrum. **70**, 2238 (1999).

[5]     M. Holt, R. Harder, R. Winarski, and V. Rose, Ann. Rev. Mater. Res. **43**, 183 (2013).

[6]     A. Pateras, M. Allain, P. Godard, L. Largeau, G. Patriarche, A. Talneau, K. Pantzas, M. Burghammer, A. A. Minkevich, and V. Chamard, Phys. Rev. B **92**, 205305 (2015).

[7]     S. O. Hruszkewycz, M. Allain, M. V. Holt, C. E. Murray, J. R. Holt, P. H. Fuoss, and V. Chamard, Nat. Mater. **16**, 244 (2016).

[8]     H. M. L. Faulkner and J. M. Rodenburg, Phys. Rev. Lett. **93**, 023903 (2004).

[9]     J. Als-Nielsen and D. McMorrow, *Elements of Modern X-ray Physics* (Wiley, Singapore, 2011).

[10]    A. Authier, *Dynamical Theory of X-Ray Diffraction* (Oxford University Press, 2001).

[11]    A. G. Shabalin, O. M. Yefanov, V. L. Nosik, V. A. Bushuev, and I. A. Vartanyants, Phys. Rev. B **96**, 064111 (2017).

[12]    W. Hu, X. Huang, and H. Yan, J. Appl. Cryst. **51**, 167 (2018).

[13]    J. A. Tilka, J. Park, Y. Ahn, A. Pateras, K. C. Sampson, D. E. Savage, J. R. Prance, C. B. Simmons, S. N. Coppersmith, M. A. Eriksson, M. G. Lagally, M. V. Holt, and P. G. Evans, J. Appl. Phys. **120**, 015304 (2016).

[14]  S. M. Durbin and G. C. Follis, Phys. Rev. B **51**, 10127 (1995).

[15]  B. W. Batterman and H. Cole, Rev. Mod. Phys. **36**, 681 (1964).

[16]  V. I. Punegov, S. I. Kolosov, and K. M. Pavlov, J. Appl. Cryst. **49**, 1190 (2016).

[17]  S. Gehrsitz, H. Sigg, N. Herres, K. Bachem, K. Kohler, and F. K. Reinhart, Phys. Rev. B **60**, 11601 (1999).

[18]  P. Delogu, P. Oliva, R. Bellazzini, A. Brez, P. L. de Ruvo, M. Minuti, M. Pinchera, G. Spandre, and A. Vincenzi, J. Instrum. **11**, 01015 (2016).

[19]  A. Ying, B. Osting, I. C. Noyan, C. E. Murray, M. Holt, and J. Maser, J. Appl. Cryst. **43**, 587 (2010).

[20]  J. Goodman, *Introduction to Fourier Optics* (McGraw-Hill, New York, 1996).

[21]  S. Nakatani and T. Takahashi, Surf. Sci. **311**, 433 (1994).

[22]  A. Pateras, J. Park, Y. Ahn, J. A. Tilka, M. V. Holt, C. Reichl, W. Wegscheider, T. A. Baart, J.-P. Dehollain, U. Mukhopadhyay, L. M. K. Vandersypen, and P. G. Evans, Nano Lett. **18**, 2780 (2018).

[23]  W. Chao, J. Kim, S. Rekawa, P. Fischer, and E. H. Anderson, Optics Exp. **17**, 17669 (2009).

[24]  J. Vila-Comamala, A. Diaz, M. Guizar-Sicairos, A. Mantion, C. M. Kewish, A. Menzel, O. Bunk, and C. David, Opt. Express **19**, 21333 (2011).

FIG. 1. (a) Crystal-thickness dependence of the angular full-width at half maximum (FWHM) of the GaAs (004) reflection in (blue) kinematical and (black) dynamical theory calculations. Dashed red and green lines indicate the thickness corresponding to the x-ray extinction depth and the Darwin width, respectively. (b) Crystal-thickness dependence of the peak reflectivity of the GaAs (004) reflection in kinematical and (black) dynamical calculations. Dashed red and orange lines indicate the thickness corresponding to the x-ray extinction depth and unity reflectivity, respectively.

FIG. 2. (a) Kinematical and (b) dynamical theory predictions of the diffraction patterns the GaAs/AlGaAs heterostructure for an incident x-ray plane wave.

FIG. 3. (a) GaAs/AlGaAs quantum well heterostructure. (b) Synchrotron x-ray nanobeam diffraction experiment.

FIG. 4. (a) Nanobeam diffraction pattern acquired at a nominal incident angle of 24.95°, matching the GaAs (004) Bragg condition. (b) Simulated diffraction pattern calculated for the same incident angle. (c) Vertically integrated intensity profiles of diffraction patterns from (a) and (b), plotted as a function of the angular difference from the center of distribution of diffracted intensity, $2\theta - 2\theta_{center}$.

FIG. 5. (a)-(d) Experimental and (e)-(h) simulated diffraction patterns corresponding to the incident angles indicated at the top right of each panel.

Pateras et *al*. Figure 1

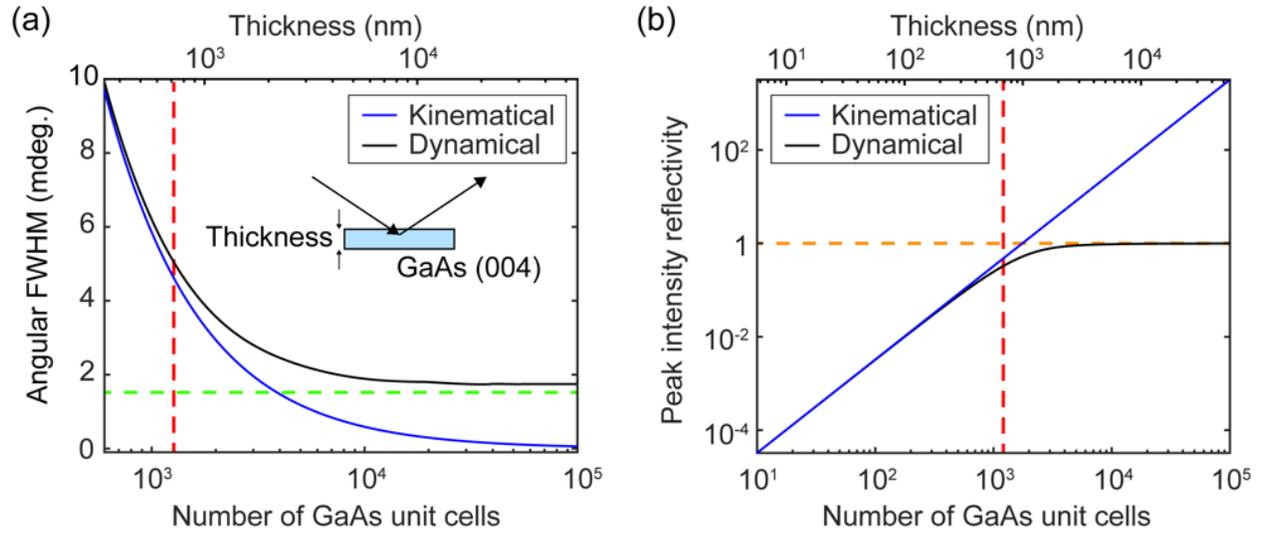

Pateras et al. Figure 2

(a) 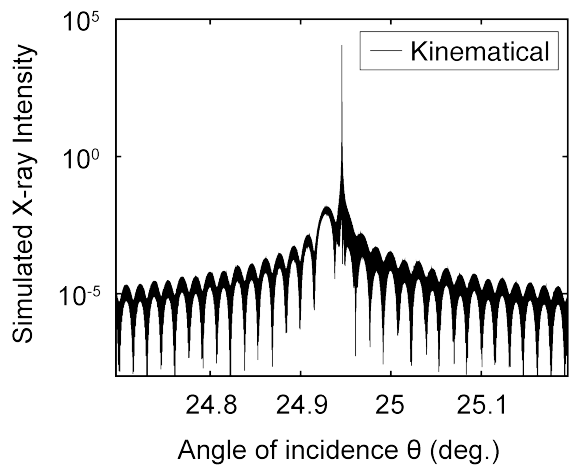

(b) 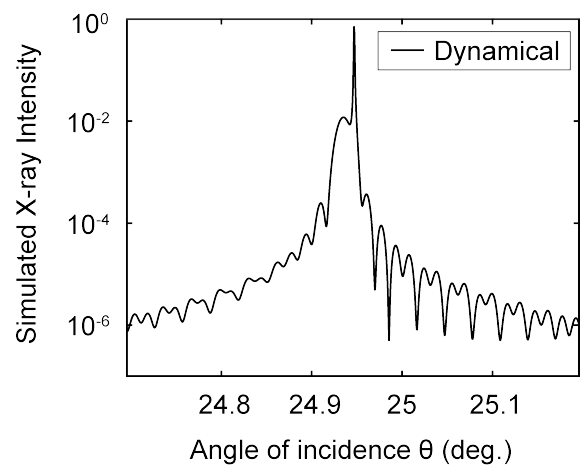

Pateras et *al.* Figure 3

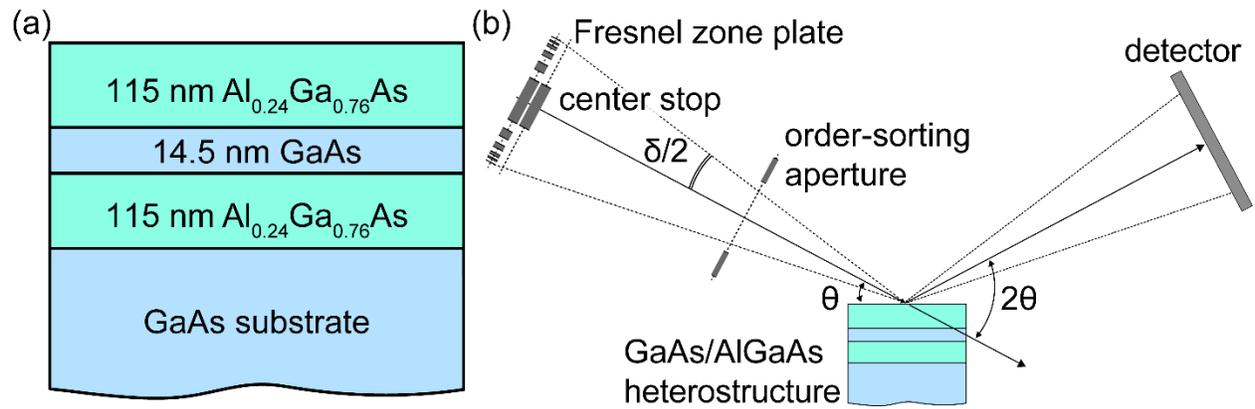

Pateras et *al.* Figure 4

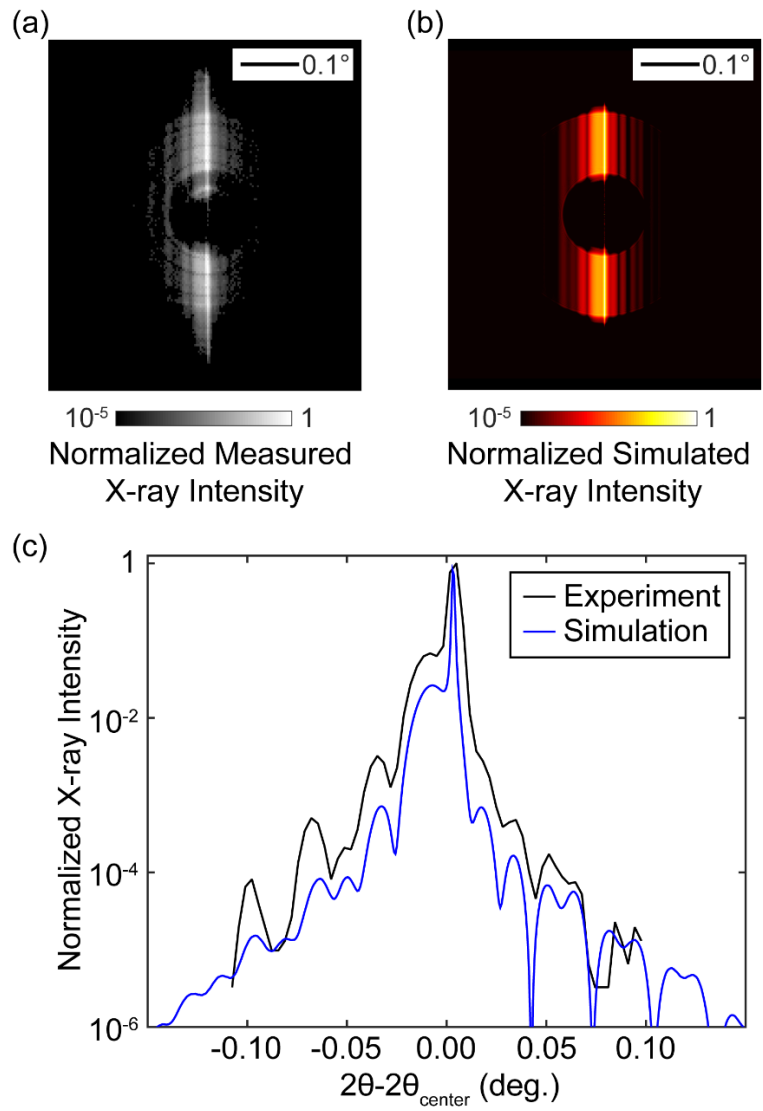

Pateras et *al.* Figure 5

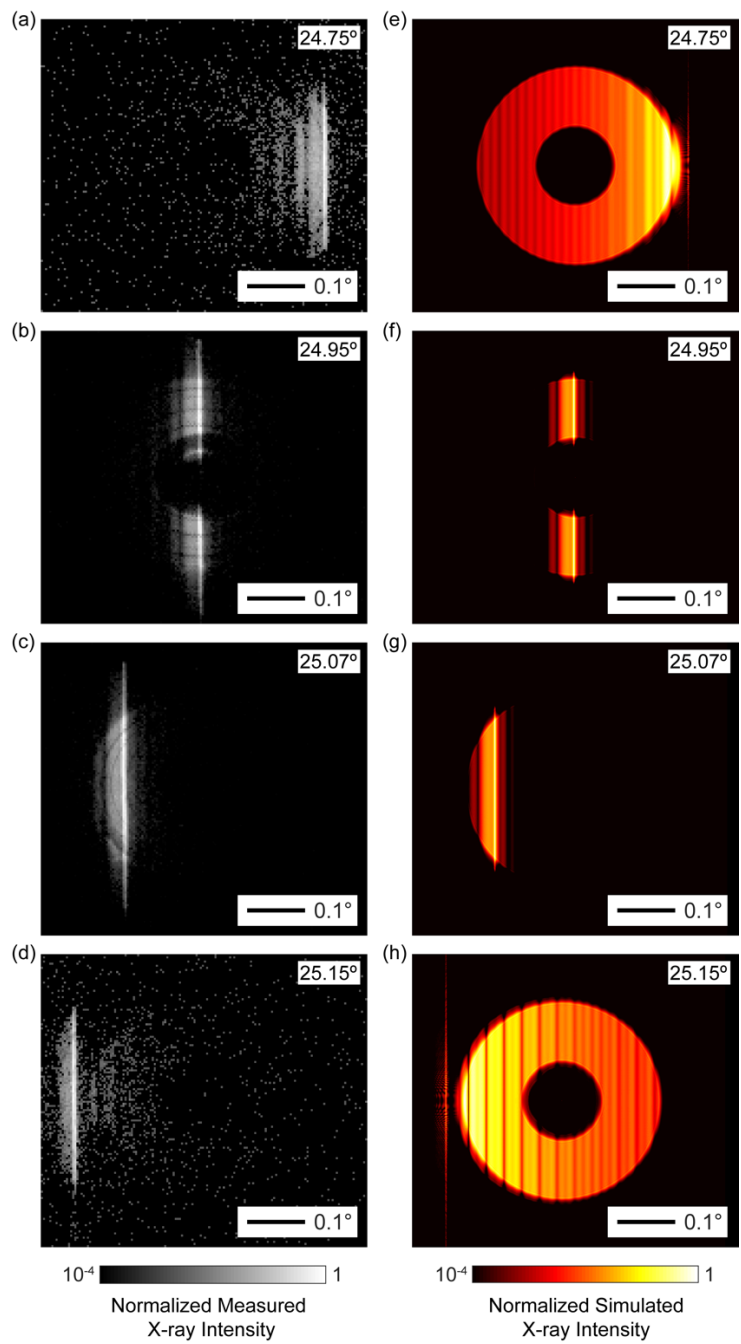